\documentclass[aps,prd,showpacs,preprintnumbers,nofootinbib,floatfix,onecolumn]{revtex4}

\usepackage{bm}
\usepackage{latexsym}
\usepackage{dcolumn}
\usepackage{amsfonts,amssymb}
\usepackage{graphicx,epsfig}
\usepackage{psfrag}
\usepackage{subfigure}

\def\beq{\begin{equation}}
\def\eeq{\end{equation}}
\def\br{\begin{eqnarray}}
\def\er{\end{eqnarray}}

\def\l{\left}
\def\r{\right} 
\def\nn{\nonumber}   
\def\eq#1{{Eq.~(\ref{#1})}}

\begin{document}
  

\title{Particle creation in a time-dependent electric field revisited}
\author{Gaurang Mahajan}
\email[]{gaurang@iucaa.ernet.in}
\affiliation{IUCAA, Post Bag 4, Ganeshkhind, Pune - 411 007, India}

\date{\today}

\reversemarginpar

\begin{abstract} 
We adopt the general formalism for analyzing evolution of gaussian states of quantized fields in time-dependent backgrounds in the Schrodinger picture (presented in detail in \cite{gm}) to study the example of a spatially uniform electric field background (in a time-dependent gauge) which is kept turned on for a finite duration of time. In particular, we study the \emph{time-dependent} particle content, defined in terms of the concept of instantaneous eigenstates, and describe how it captures the time evolution of the quantized field modes. The actual particle creation process occurs over a relatively short interval in time, and the particle content saturates rather quickly. We also compare the \emph{power spectrum} of the field modes, computed in the asymptotic limit, with the corresponding situation in a cosmological de Sitter background. Particle creation under the influence of a spiked electric field localized in time, as a particular limiting case of the above general model, is also considered.
\end{abstract}

\maketitle


\section{Introduction}

The Schwinger effect, involving pair creation by a classical electromagnetic field out of the vacuum, is a well-known representative example of quantum field theory in an external classical background~\cite{schwinger,txts1}. The usual treatment, in the case of a spatially uniform and constant (in time) electric field, proceeds by evaluating the gauge-invariant effective action which is directly related with the transition amplitude between asymptotically defined \emph{in-out} vacuum states~\cite{schwinger,ef1}. The real part of the effective Lagrangian describes vacuum polarization by the electric background, and can be added to the classical electromagnetic Lagrangian to compute the modified equations of motion for the background incorporating back-reaction at the semiclassical level. The imaginary part encodes information about particle creation, or more precisely, provides a direct measure of the asymptotic particle content in the \emph{in} vacuum with respect to the \emph{out} vacuum state. An alternative route to deriving this result is based on a particular choice for the gauge in which the vector potential is time-dependent. The field modes can be treated as a bunch of quantized time-dependent oscillators, and sensible vacuum states can be defined in the asymptotic regions~\cite{ef1,ef2}. These states are however in-equivalent due to the non-trivial time dependence of the vector potential -- though the electric field is constant -- with the consequence that a positive frequency mode defined at early times evolves into a combination of positive and negative frequency modes with respect to late times. This allows one to compute the Bogolyubov coefficients for the mixing, and hence the mean number of particles in any given mode as seen by an observer at late times.  

Both the above approaches however involve only the asymptotically defined vacuum states, and do not really provide a direct picture of the time evolution of the quantized field. Moreover, this approach to account for back-reaction is conceptually somewhat unsatisfactory from the point of view of causality, since one is, in effect, getting an effective Lagrangian which depends implicitly on the \emph{out} vacuum state. Further, one might have thought that physically realistic electromagnetic fields would exist for finite duration and it makes more sense to have a situation where the field is kept switched on for a finite period of time. In fact a constant, eternally existing electric field would be a rather unphysical notion. This is because if the field is causing particles to be created continuously, the produced pairs would `short' the externally applied field, and cause it to progressively decay over time. An eternal electric field would thus be unstable and cannot really be maintained when particle creation is taken into account. But if one considers the case of a field which exists for only a finite duration, it will naturally be expected to break the time translation symmetry of the eternal field case, introducing a finite window in time for quantum effects to come into play and leading to a time-dependent picture of the particle creation process. One can make an analogy between these two cases and those of (i) an eternal black hole and (ii) a black hole formed from gravitational collapse.  In the case of (i), for a quantized field in the Hartle-Hawking state~\cite{txts2}, an observer at spatial infinity would not see a flux of particles because of the time translation symmetry of the gravitational background. When one considers a collapsing scenario, however, one is introducing a time dependence into the gravitational field, and in such a background, there would be an outgoing flux of particles (in the Unruh vacuum), carrying away energy which would cause the black hole to evaporate over time. 

In~\cite{gm} we adopted a physically reasonable, time-dependent approach to analyzing field modes in the Schrodinger picture, and used an `instantaneous' definition of particle. The Schwinger effect is amenable to this treatment, as we analyzed in considerable detail for an eternal electric field. But such a scenario is time translation-invariant, and we would like to adopt the same approach and repeat the analysis for a spatially uniform but time-dependent electric field. Here we consider the case of a field which exists for a \emph{finite} duration of time $T$, tracking the behavior of the time-dependent particle number and interpreting it appropriately. For an electric field of magnitude $E$, the relevant time scale (in natural units) for quantum effects is provided by $\tau \sim 1/\sqrt{qE}$. On physical grounds, the results for the late-time particle content etc. for the eternal electric field scenario would be expected to be recovered when $T$ is made very large, i.e. $T/\tau \gg 1$, and we would like to verify that this is indeed the case. We will also attempt to connect up the instantaneous particle content with the power spectrum of the quantum field and indicate the distinction between the two. For contrast, we also discuss the other extreme of an electric field sharply localized in time modeled by a Dirac delta function. The question of having a notion of classicality of the evolving quantum state, which was extensively dealt with in~\cite{gm}, will also be touched upon, because it has potential implications for the study of back-reaction at the semiclassical level and for the validity of a semiclassical analysis.

This paper is structured as follows. In Section~\ref{sec:formalism} we sketch an outline of the Schrodinger picture formalism which was elaborated on in \cite{gm}, to analyze a quantized field, treated as a bunch of oscillators, in the presence of a time-dependent vector potential. In Section~\ref{sec:analysis} the analysis of time evolution of any single fourier mode of the quantized field in the presence of a background electric field is carried out for the two specific cases mentioned above: a field which exists for a finite period of time (and is maintained constant over that period), and a spiked $\delta$-function electric field. Finally, we conclude with a recapitulation in Section~\ref{sec:conc}.
In what follows, we shall set $\hbar=c=1$.    


\section{Formalism}  \label{sec:formalism}

We start with a complex scalar field $\phi$ which is quantized in a spatially uniform electric field background. The electric field ${\bf E}(t)$ is assumed to be produced by a (uniform) current density ${\bf J} (t)$, and can be described in a time-dependent gauge corresponding to $A^i \equiv (0,0,0,A(t))$. We are also assuming that the magnetic field is zero everywhere, and so is the electric charge. Then, from Maxwell's equations, it follows that $J(t)=(1/4\pi)\ddot{A}(t)$.

The action for the scalar field in this background is expressible as~\cite{schwinger,ef1,ef2}
\beq
{\cal A}[\phi] = \frac{1}{2}\int {d^{4}x}~(\partial_{\mu} + iqA_{\mu})\phi~(\partial^{\mu} - iqA^{\mu})\phi^{*} \nonumber\\
\eeq
\beq
= \frac{1}{2} \int {d^3{\bf{k}}} \int dt \left(\vert \dot q _{\bf{k}}\vert ^2 - (k_{\perp}^2 + m^2 + (k_{z} - q A(t))^2 )\vert q_{\bf{k}}\vert ^2\right)    \label{ef_actn}
\eeq
with ${\bf k}\cdot {\bf k} =k^2_{\perp}+k^2_{z}$ and ${\bf k}_\perp \cdot {\bf E} = 0$. The field theory thus reduces to quantum mechanics of a bunch of uncoupled oscillators each with unit mass and a time-dependent frequency given by
\beq
\omega_{\bf k}(t)=\sqrt{k_{\perp}^2 + m^2 + (k_{z} - q A(t))^2}.
\eeq
(Since the scalar field is complex-valued, each mode $\bf{k}$ is associated with two degrees of freedom, corresponding to the real and imaginary parts of $q_{\bf{k}}$.)

Having cast the problem in a time-dependent form, we now briefly outline the Schrodinger picture formalism (based on~\cite{gm}) which will be adopted here. The starting point is the wave function, chosen to be a \emph{form-invariant} gaussian state (motivated by the requirement of having a state that can be interpreted as the ground state of the oscillator in a time-independent scenario) parameterized by the complex variable $z_{\bf k}(t)$:
\beq
\psi (q_{\bf k},t) = N_{\bf k}(t) \exp \l[ - \frac{\omega_{\bf k}}{2} \l(\frac{1-z_{\bf k}}{1+z_{\bf k}} \r) q_{\bf k}^{2} \r].    \label{gwfn}
\eeq
The time dependent Schrodinger equation reduces to a first order (but nonlinear) differential equation for $z_{\bf k}(t)$:
\beq
\dot z_{\bf k} + 2 i \omega_{\bf k} z_{\bf k} + \frac{1}{2} \frac{\dot \omega_{\bf k}}{\omega_{\bf k}} (z_{\bf k}^{2} - 1) = 0.  \label{eqz}
\eeq
The problem of determining the quantum evolution thus boils down to directly solving for $z_{\bf k}$ (with an appropriate choice of initial condition). The dimensionless adiabaticity parameter $\dot{\omega_{\bf k}}/\omega_{\bf k}^2 \equiv \epsilon_{\bf k}(t)$, provides the relevant measure to characterize the nature of the time evolution; $\vert \epsilon_{\bf k}\vert \ll 1$ corresponds to the adiabatic limit, in which region a reasonable definition of vacuum state is possible. 

In~\cite{gm}, we introduced a time-dependent notion of particle. This is done by comparing the quantum state moment by moment  with the complete set of \emph{instantaneous} eigenstates of the oscillator defined at every instant. This provides a number distribution of `quanta', and can be used to compute the mean, which provides a rather general notion of particle content. It is given by
\beq
\langle n_{\bf k} \rangle = \frac{|z_{\bf k}|^2}{1-|z_{\bf k}|^2} \quad;\quad {\cal E}_{\bf k} = \l( \langle n_{\bf k} \rangle + \frac{1}{2} \r) \omega_{\bf k}(t) 
\eeq 
and is thus directly related to the expectation value of the time-dependent Hamiltonian. Once $z_{\bf k}(t)$ is specified, the evolution of this particle number can be determined.

It may be noted that the mean particle number as well as the higher moments of the number probability distribution are functions of only the magnitude of $z_{\bf k}$, and do not encode information about its phase. The phase can be fixed by considering the variance of $q_{\bf k}$, generally called the power spectrum, which is the fourier transform of the two-point correlation function of the field in coordinate space. The expression for the power in the ${\bf k}$th mode has the form
\beq
\langle q_{\bf k}^{2} \rangle = \frac{\l ( 2 \langle n_{\bf k} \rangle + 1 \r)}{2 \omega_{\bf k}} + \frac{(\langle n_{\bf k} \rangle +1)}{ \omega_{\bf k} } \textrm{Re}(z_{\bf k}).  \label{ps_ef}
\eeq 
which makes it clear that the power spectrum is not, in general, \emph{completely} specifiable in terms of the mean number of created particles. This point will be taken up later and a comparison with the case of an inflationary cosmological model made.

Returning to the quantum state in \eq{gwfn}, we can discuss another question, that of identifying a suitable criterion for classicality of the state. The Wigner function~\cite{gm,paddy,wig} provides one reasonable approach. In~\cite{gm}, we looked at two features as possible indicators of classical-like behavior: peaking of the Wigner distribution in phase space and correlation (in the sense of separability of the Wigner function in dependence on $q$ and $p$). The general conclusion drawn was that the former is not an unambiguous indicator of classical behavior; although indicative of a correspondence between position and momentum, it can happen even in a near-vacuum state. The classicality parameter defined as ${\cal S}_{\bf k}=\langle q_{\bf k} p_{\bf k}\rangle /\sqrt{1 + \langle q_{\bf k} p_{\bf k}\rangle ^2}$, on the other hand, is directly related to the mean of $q_{\bf k} p_{\bf k}$ and shows a close correspondence with particle production, being zero for an instantaneous ground state but saturating at unity when the mean particle number diverges. We will consider its variation too in the analysis which follows.

We now move on to describing the setting in which the analysis has been carried out, that of a spatially uniform but time-dependent electric field.


\section{Analysis}  \label{sec:analysis}        

\subsection{A spatially constant electric field with a finite extent in time} \label{box}

We consider a situation in which the electric field is kept switched on for a finite period of time $T$, from $t=-T/2$ to $t=T/2$. The time-dependent vector potential is chosen to have the following form:
\br
A(t) &=& ET/2 \quad (t < T/2) \nn \\
     &=& -Et   \qquad (\vert t \vert < T/2) \nn \\
     &=& -ET/2 \quad (t > T/2).                \label{sc_ef}
\er
Such a field would be produced by a spatially uniform current density $J(t)$ which remains zero at all times except at $t=\pm T/2$, being made up of two $\delta$-function spikes separated by a time interval $T$.
 
Making the transformation to dimensionless parameters $x=\sqrt{qE}t$, $k_z/\sqrt{qE}=\xi_{\bf k}$ and $(k_\perp ^2 + m^2)/qE=\lambda_{\bf k}$, the frequency function corresponding to \eq{sc_ef} can be written as
\br
\omega_{\bf k}(x) &=& \sqrt{\lambda_{\bf k} + (\xi_{\bf k} + x)^2} \qquad (|x|\leq x_0 \equiv \sqrt{qE}T/2) \nn \\
&=& \sqrt{\lambda_{\bf k} + (\xi_{\bf k} \mp x_0)^2} \equiv \omega_{\bf k} ^{\mp} \qquad ( x < - x_0 , x > x_0).
\er 
The oscillator equation for the variable $z_{\bf k}(x)$ can be solved analytically for this choice of frequency in terms of the parabolic cylinder function~\cite{table}; assuming that the field is in the vacuum state initially (for $x < -x_0$), the solution for $z_{\bf k}$ (which is continuous across the interval where $E \neq 0$) is given by 
\beq
z_{\bf k}(x) = \frac{f'_{\bf k}(x) - i \omega_{\bf k} (x)f_{\bf k}(x)}{f'_{\bf k}(x) + i \omega_{\bf k} (x)f_{\bf k}(x)}
\eeq
for $\vert x \vert \leq x_0$, where
\beq
f_{\bf k}(x) = D_{-1/2 - i\lambda_{\bf k}/2}[-(1+i)(\xi_{\bf k} + x)] + \gamma_{\bf k} (D_{-1/2 - i\lambda_{\bf k}/2} [-(1+i)(\xi_{\bf k} + x)])^*
\eeq
and
\beq
\gamma_{\bf k} = \frac{D'_{-1/2 - i\lambda_{\bf k}/2}[-(1+i)(\xi_{\bf k} - x_0)] - i \omega_{\bf k}^{-} D_{-1/2 - i\lambda_{\bf k}/2} [-(1+i)(\xi_{\bf k} - x_0)]}{-D'^* _{-1/2 - i\lambda_{\bf k}/2}[-(1+i)(\xi_{\bf k} - x_0)] +  i \omega_{\bf k}^{-} D^* _{-1/2 - i\lambda_{\bf k}/2} [-(1+i)(\xi_{\bf k} - x_0)]}.
\eeq 
For $x \geq x_0$, the solution is
\beq
z_{\bf k}(x) = \l( \frac{f'_{\bf k}(x_0) - i \omega _{\bf k}(x_0)f_{\bf k}(x_0)}{f'_{\bf k}(x_0) + i \omega _{\bf k}(x_0)f_{\bf k}(x_0) } \r) e^{-2 i \omega_{\bf k}^{+} (x - x_0)}.  \label{n_scef}
\eeq
From the above expression, it is immediately obvious that $z_{\bf k} \neq 0$ for $x > x_0$, indicating a non-zero particle content in the quantum state which remains constant once the electric field has been switched off, since all the time dependence is in the phase of $z_{\bf k}$. We are, however, particularly interested in the \emph{time dependence} of the particle creation process. (Although the electric field is constant within the time interval $T$, the particle content is expected to exhibit time dependence since we have cast the problem in a time-dependent form by a particular gauge choice.) Tracking the variation of the mean number $\langle n_{\bf k} \rangle$ can provide an idea about this.

\begin{figure}[h]
\begin{center}
\subfigure[ ]{\label{n3}\includegraphics[width=5cm,angle=0.0]{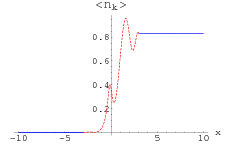}}
\hspace{0.3cm}
\subfigure[ ]{\label{n5}\includegraphics[width=5cm,angle=0.0]{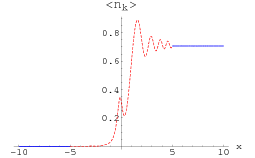}}
\hspace{0.3cm}
\subfigure[ ]{\label{n8}\includegraphics[width=5cm,angle=0.0]{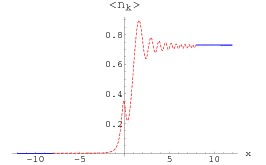}}
\end{center}
\caption{Variation of the mean particle number $\langle n_{\bf k} \rangle$ (for $\xi_{\bf k}=0.1$ and $\lambda_{\bf k} = 0.1$) as a function of the dimensionless time parameter $x$ for $x_0 =3, 5$ and 8. The dotted portion corresponds to the interval when the field is present. From a comparison of the plots, the `final' mean particle number (constant for $x \geq x_0$) evidently seems to depend on how long the field is present (at least for the range of values of $x_0$ considered here); but a closer look suggests that the particle content might saturate eventually.}
\label{n-vs-x}
\end{figure}

\begin{figure}[h]
\begin{center}
\label{n_x0}\includegraphics[width=10cm,angle=0.0]{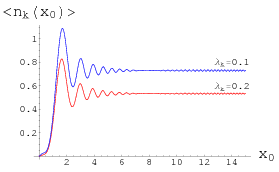}
\end{center}
\caption{Plot of the mean particle number $\langle n_{\bf k} \rangle$ at $x=x_0$ as a function of $x_0$ for $\xi_{\bf k}=0.1$ and $\lambda_{\bf k} = 0.1$ (blue) and 0.2 (red). $x_0$ is proportional to the interval of time over which the electric field is present. This confirms what the earlier plots suggested. The saturation beyond a point indicates that in the picture adopted here, for large enough $x_0$, the final particle content becomes independent of how long the field is kept switched on.}
\label{n-vs-x0}
\end{figure}

Fig.~\ref{n-vs-x} shows the variation of $\langle n_{\bf k} \rangle$ as a function of the variable $x$ for the cases $x_0 = $ 3, 5 and 8. The red portion corresponds to the interval over which the electric field is kept switched on. As expected, the particle number rises from zero to a finite value at $x=x_0$ and remains constant thereafter, once the field is turned off. This rise is however not monotonic, and is characterized by oscillations during the phase when the field is present. This is of course at odds, to some extent, with one's intuitive notion of a `particle', once produced, remaining produced; an appropriate interpretation of $\langle n_{\bf k} \rangle$ in the intermediate phase is closer to a measure of quantum fluctuations of the scalar field rather than a measure of physical particles. If the time scale of the electric field is increased, it is evident from a comparison of the plots that these oscillations get progressively (and rather quickly) suppressed, and the particle number very soon reaches a saturation. In fact, the particle creation is seen to really happen only over a rather short interval. For sufficiently large values of $x_0$, the value of $\langle n_{\bf k} \rangle$ at $x=x_0$ is more or less independent of $x_0$ (and is equal to the asymptotic value for an eternal electric field, usually determined from the effective action or using the method of Bogolyubov coefficients~\cite{ef1}). This can be gleaned from the plot in Fig.~\ref{n-vs-x0} of $\langle n_{\bf k} \rangle (x_0)$ as a function of $x_0$. To restate the obvious, an electric field which is kept turned on long enough gives essentially the same result for the late-time particle content as the standard constant field background.

\begin{figure}[h]
\begin{center}
\subfigure[ ]{\label{S3}\includegraphics[width=5cm,angle=0.0]{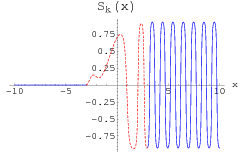}}
\hspace{0.3cm}
\subfigure[ ]{\label{S5}\includegraphics[width=5cm,angle=0.0]{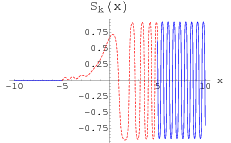}}
\hspace{0.3cm}
\subfigure[ ]{\label{S8}\includegraphics[width=5cm,angle=0.0]{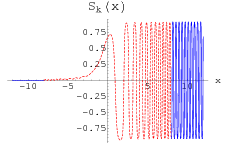}}
\end{center}
\caption{Time evolution of the classicality parameter $\cal{S}_{\bf k}$ (for $\xi_{\bf k}=0.1$ and $\lambda_{\bf k} = 0.1$) for the cases $x_0 = 3, 5$ and 8. The correlatedness grows, and on an average is larger at late times in comparison with early times (being zero there). The oscillatory behavior reflects the underlying finite and steady-state particle content attained by the quantum state.}
\label{S-vs-x}
\end{figure}

As for the classicality parameter, shown in Fig.~\ref{S-vs-x}, it starts from zero corresponding to the initial ground state, increases over a short interval of time and then settles down into a steady oscillatory pattern of constant amplitude. One can interpret this as a growth in the phase space correlation over time starting from an initially uncorrelated vacuum state, although $\langle q_{\bf k} p_{\bf k} \rangle$ (not its magnitude) is oscillatorily zero at late times.

From the previous plots, it can be seen that for large enough $\sqrt{qE}T$, the behavior for $t > T$ closely resembles that in the time-independent case. We shall now take the $\sqrt{qE}T \sim x_0 \to \infty$ limit, and look at the behavior for $ x \to \infty $. The power spectrum in this limit~\cite{gm} is given by the expression
\beq
\langle q_{\bf k}^{2} \rangle \approx \l(\frac{ 2 e^{-\pi \lambda_{\bf k}}  + 1 }{2 x}\r) + \l(\frac{e^{-\pi \lambda_{\bf k}} +1}{ x }\r) \textrm{Re} \l(\frac{e^{-i \l(x^{2} + \lambda_{\bf k}\ln x \r)}}{{\cal R}_{\bf{k}}} \r)  
\eeq 
which can be written as
\beq
\langle q_{\bf k}^{2} \rangle \approx \frac{1}{x}\l(e^{-\pi \lambda_{\bf k}}  + \frac{1}{2}+ \sqrt{e^{-\pi \lambda_{\bf k}}\l( e^{-\pi \lambda_{\bf k}} + 1 \r)} \cos \l( x^{2} + \lambda_{\bf k}\ln x + r_{\bf{k}} \r) \r)  \label{psef}
\eeq
where 
\beq
{\cal R}_{\bf{k}} \equiv \vert {\cal R}_{\bf{k}} \vert \exp(ir_{\bf{k}}) = -\sqrt{2 \pi} i \frac{ \exp \l(\frac{\pi}{4} \lambda_{\bf k}\r)}{\Gamma \l( \frac{1}{2} + i \frac{\lambda_{\bf k}}{2} \r)}.
\eeq
 If the third term in the brackets in \eq{psef}, which is rapidly oscillating, is dropped, then  the variance at the leading order is just 
\beq
\langle q_{\bf k}^{2} \rangle \approx \frac{1}{x}\l( \langle n_{\bf k}\rangle + \frac{1}{2} \r)
\eeq
 and the power in the large time limit can be related to the asymptotic mean particle number. This connection is not quite precise though, since it requires our ignoring the additional oscillatory term which amounts to ignoring the phase information contained in $z_{\bf k}$.  

We can go a bit further and try computing the spatial two-point correlation function. In the large $x$ limit, if the oscillatory term in \eq{psef} is ignored, the fourier transform of the first term can be found, and one has
\beq
\langle \phi({\bf r})\phi(0) \rangle \quad \simeq \quad \frac{1}{2}\delta({\bf r}) + \frac{qE}{x}e^{-\pi\frac{m^2}{qE} - \frac{qE}{4 \pi} {\bf r}_{\perp}^2 } \delta(z)
\eeq
where $\bf{r}_\perp \cdot \bf{E} = 0$. The second term in the above expression suggests that on any 2D plane normal to the direction of the applied electric field, the correlation between any two points falls off for distances beyond $L_c \sim 1/\sqrt{qE}$; the spatial two-point function, in the late time limit, thus picks out the characteristic length scale for correlations.

This behavior of the power spectrum may be compared with the corresponding evolution of a massless real scalar field in an expanding FRW cosmological background~\cite{paddy,txts2}; the general expression for the power happens to be~\cite{gm}:
\beq
\langle q_k^{2} \rangle = \frac{\l ( 2 \langle n_k \rangle + 1 \r)}{2 k a^2} + \frac{(\langle n_k \rangle +1)}{ k a^2 }\textrm{Re}(z_k)  
\eeq
$a$ being the scale factor and $k = \vert {\bf k} \vert$ . At super Hubble scales, any mode evolves highly non-adiabatically, so $z_k \to 1$ and $\langle n_k \rangle \gg 1$, giving $\langle q_k^{2} \rangle \approx 2 \langle n_k \rangle /(k a^2)$, and the power in any given fourier mode provides a direct measure of its particle content. In particular, for the de Sitter model,
\beq
\langle q_k^{2} \rangle = \frac{H^2}{2 k^3}\l( 1 + \frac{k^2}{a^2 H^2} \r). 
\eeq
At late times ($a \to \infty$), $\langle q_k^{2} \rangle$ becomes \emph{constant} and proportional to $k^{-3}$, which corresponds to a scale-invariant spectrum. The scale invariance, moreover, implies that the spatial two-point function here does not pick out a well-defined correlation length scale, and this is in contrast with the electric field case.   

\subsection{An electric field narrowly localized in time}

In the foregoing analysis we considered an electric field of finite strength which is kept switched on for a finite duration, before shifting attention to the limiting case of large timescales (keeping the magnitude of the field fixed). It would be of interest to also look at the other extreme, that of a field sharply peaked in time. 

Let us consider the case of a spatially constant electric field modeled by a Dirac delta function in time: $E(t) = \alpha \delta (t)$. This can be described by the vector potential $A(t) = - \alpha \theta(t)$, where $\theta(t)$ denotes the Heaviside theta function. The potential and hence the frequency $\omega_{\bf k} (t)$ have a discontinuity at $t=0$, and it is this step which will create particles.

\begin{figure}[h]
\begin{center}
\includegraphics[width=8cm,angle=0.0]{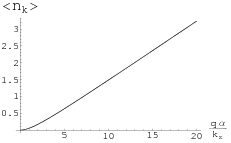}
\end{center}
\caption{The (constant) mean number of particles for $t > 0$ created by the delta function electric field at $t=0$ as a function of $q \alpha / k_z$. The plot corresponds to a mode with $k_{\perp}^2 + m^2 = k_z ^2$. The particle content monotonically increases with $q\alpha$, diverging in the $q\alpha \to \infty$ limit. }
\label{n_df}
\end{figure}

In order to compute the particle creation at the jump, we will work with the mode variable $f_{\bf k}$, which satisfies the standard oscillator equation:
\beq
\ddot{f_{\bf k}} +  \omega_{\bf k}^2 (t) f_{\bf k} = 0.
\eeq 
$f_{\bf k}$ is related to the variable $z_{\bf k}$ by the relation
\beq
z_{\bf k} = \frac{i \omega_{\bf k} - \dot{f_{\bf k}}/f_{\bf k} }{i \omega_{\bf k} + \dot{f_{\bf k}}/f_{\bf k}}.
\eeq
In this case, the oscillator frequency makes a sharp transition from $\omega_{\bf k}^{i} = \sqrt{k_\perp ^2 + m^2 + k_z ^2}$ for $t<0$ to $\omega_{\bf k}^{f} = \sqrt{k_\perp ^2 + m^2 + (k_z + q \alpha)^2}$ for $t>0$.  If it is assumed that the mode starts off in the vacuum state at some $t < 0$, which corresponds to setting $z_{\bf k}(t<0) = 0$, then it follows that $ f_{\bf k}  \propto \exp(i \omega_{\bf k}^{i} t)$. Imposing continuity of $f_{\bf k}$ and $\dot{f_{\bf k} }$ at the jump at $t=0$, one can then evaluate the mode function for $t>0$. This turns out to be given by
\beq
f_{\bf k} (t>0)  =  \frac{1}{\sqrt{2 \omega_{\bf k}^{f}}} \l( A_{\bf k} e^{i \omega_{\bf k}^{f} t} +  B_{\bf k} e^{-i \omega_{\bf k}^{f} t} \r)
\eeq
where 
\beq
\frac{B_{\bf k}}{A_{\bf k}} = \l( \frac{\omega_{\bf k}^{f} - \omega_{\bf k}^{i}}{\omega_{\bf k}^{f} + \omega_{\bf k}^{i}}  \r ).
\eeq
Using this, the form of $z_{\bf k}$ for $t>0$ is determined to be
\beq
z_{\bf k}(t) = \l( \frac{\omega_{\bf k}^{f} - \omega_{\bf k}^{i}}{\omega_{\bf k}^{f} + \omega_{\bf k}^{i}}  \r ) e^{-2 i \omega_{\bf k}^f t},
\eeq
which yields the following expression for the mean number of particles created at the jump:
\beq
\langle  n_{\bf k} \rangle = \frac{|z_{\bf k}|^2}{1-|z_{\bf k}|^2}  = \frac{1}{4}\l( \sqrt{\frac{\omega_{\bf k}^{f}}{\omega_{\bf k}^{i}}} -  \sqrt {\frac{\omega_{\bf k}^{i}}{\omega_{\bf k}^{f}}} \r)^2.   \label{n_dfn}
\eeq
(One can alternatively impose continuity of the wavefunction given by \eq{gwfn} at $t=0$ to directly obtain the jump in $z_{\bf k}$.) From this expression, it is clear that in the limit of large impulse ($q \alpha \to \infty$), we have $\omega_{\bf k}^{f}/\omega_{\bf k}^{i} \to \infty$ and the mean diverges. The variation of the particle content as a function of the spike strength for a particular mode is shown in Fig.(\ref{n_df}). The mean particle production is clearly a monotonically increasing function of $q\alpha$. 

It may be noted that when working directly with the wavefunction (or with $f_{\bf k}$), one does not need to choose a value for $\omega_{\bf k}(0)$ in deriving the result for $z_{\bf k}(t)$, and it can be left unspecified. If one takes the alternative route of working with the evolution equation (\ref{eqz}) instead, then the nature of the differential equation brings in a dependence on $\omega_{\bf k}(0)$. But the requirement of consistency with the earlier method uniquely fixes its value. This may be seen as follows: Let us start with the \eq{eqz}. This equation can be integrated over a small interval $-\epsilon < t < \epsilon$ around $t=0$ followed by taking the $\epsilon \to 0$ limit. The second term in \eq{eqz} would give no contribution in this limit, but the presence of the delta function in the $\dot{\omega}/\omega$ term would give a discontinuity in the value of $z_{\bf k}$ across the jump, which is related to its value at $t=0$:
\beq
z_{\bf k} ( t \to 0^{+} ) = C_{\bf k} \l[ 1 - z_{\bf k} ^2 (0)  \r]     \label{eq:z_df}
\eeq
where
\beq
C_{\bf k} = \frac{q \alpha \left( k_z + q \alpha \theta(0) \right) }{2 \left[ k_\perp ^2 + m^2 
+ \left( k_z + q \alpha \theta(0) \right)^2 \right] } \equiv \frac{q \alpha}{2 \omega_{\bf k}^2 (0) } \left( k_z + q \alpha \theta(0) \right).
\eeq
In the above relation, $z_{\bf k}(0)$ and $\theta(0)$ are unspecified. In order for this relation to give the `correct' expression for the average particle number (i.e. one which matches with that derived earlier by imposing continuity of the wavefunction at the jump), one needs to externally specify a particular value for $z_{\bf k} (0)$, which turns out to be given by
\beq
z_{\bf k} (0) = \sqrt{1 - \frac{1}{C_{\bf k}}\l(\frac{\omega_{\bf k}^{f} - \omega_{\bf k}^{i} }{\omega_{\bf k}^{f} + \omega_{\bf k}^{i}} \r)}.   \label{z_0}
\eeq
With this choice for $z_{\bf k} (0)$, \eq{eq:z_df} gives the same expression as (\ref{n_dfn}) for the mean particle number. Of course, $\theta(0)$ is still arbitrary here, but if we equate the RHS in \eq{z_0} with the expression for $z_{\bf k}(0)$ obtained directly from the continuity of $\dot{f_{\bf k}}/f_{\bf k}$ at $t=0$, then the resulting consistency condition also fixes the value of $\theta(0)$ implicitly through the following relation: 
\beq
2 \alpha\omega_{\bf k}^{i} \l(\frac{\omega_{\bf k}^{f} - \omega_{\bf k}^{i} }{\omega_{\bf k}^{f} + \omega_{\bf k}^{i}} \r) \l( k_z + q \alpha \theta(0) \r) = \omega_{\bf k}(0) \l(\omega_{\bf k}(0) + \omega_{\bf k}^{i} \r)^2.   \label{theta_0}
\eeq
Thus, if one works with the evolution equation for $z_{\bf k}$, then by specifying the value of $\omega_{\bf k}(0)$ through the relation (\ref{theta_0}) and the value for $z_{\bf k} (0)$ as given by \eq{z_0}, the form of $z_{\bf k}(t)$ gets completely fixed and identical to the result which was derived earlier by working with the variable $f_{\bf k}$.
  
It may also be mentioned that, from the fact that the magnitude of $z_{\bf k}$ is a constant for $t >0$, the trajectory of the system in the complex $z_{\bf k}$ plane will be a circle for $t > 0$. This in turn corresponds to a finite and oscillatory phase space average of $q_{\bf k} p_{\bf k}$, so compared with the uncorrelated vacuum state prior to the application of the peaked electric field at $t = 0$, the field modes for $t>0$ can be regarded as more classical in this sense. This feature is, of course, also shared in common by the cases we have considered earlier in Section~\ref{box}.  


\section{Recapitulation}  \label{sec:conc}

Our analysis of a quantized scalar field in the background of a varying electric field 
(i.e. one which exists for a finite duration) viewed in a time-dependent gauge provides 
a nice way to visualize the time evolution of the field modes, and not just their asymptotic 
behavior. The concept of a time-dependent mean particle number that was adopted here goes over to the standard definition based on adiabatically definable vacuum states in the static \emph{in} and \emph{out} regions where no field is present, but displays oscillatory variation in a region where the evolution deviates from adiabaticity, presumably reflecting the transition in the field from quantum fluctuations to physical particles. The particle creation process in this picture is thus not steady, or even monotonic. 

For sufficiently large $T$, it is found that the particle content very nearly reaches a saturation beyond a finite time scale (while the electric field is still present), and the value of this steady state particle number is independent of the duration $T$ over which the field exists. Although the 
time dependence of the particle production here arises as a consequence of our choice of a time-dependent 
gauge to describe the electric field in, the effective Lagrangian which can be computed using this 
quantity would be gauge invariant. 

When the $\sqrt{qE} T \to \infty$ limit is taken, the late time behavior of the power spectrum of the quantum fluctuations and the corresponding two-point function indicate the setting up of field correlations over length scales of order $\lesssim 1/\sqrt{qE}$. The behavior of the 
time-dependent classicality parameter, encapsulating one aspect of the phase space evolution 
of the field modes, shows that the quantum state grows more correlated over time in 
comparison with the initial vacuum state it starts off in, and can be interpreted as turning more classical, in this sense, as particle production progresses. 

For contrast we have also looked at the case of a sharply localized electric field, which in some sense can be thought of as the $T \to 0$ limit of the earlier model with the impulse $qET$ held fixed. The mean number of particles produced per mode can be computed by imposing continuity of the wavefunction at the `jump' in the value of the frequency $\omega_{\bf k}$. It is found to monotonically increase with the spike strength, i.e. the impulse $q \alpha$, and diverges when the $q \alpha \to \infty$ limit is taken.   
\newline
\newline
\newline
{\bf Acknowledgments}: The author wishes to thank Prof. T. Padmanabhan for useful discussions and comments on various aspects of this work. The author is supported by a Fellowship from the Council of Scientific \& Industrial Research, India.


\end{document}